# Emergency Department Revisit Quality Review Screening: Exploring Human Decision-Making and Artificial Intelligence Support

**Authors (with numeric identifier for affiliation, city, state in parentheses)**: Jonathan A. Handler, MD (1, 11), Marlene I. Robles-Granda, BSc, MSc (2), Jacob E. Mefford, PA-C (3), Jeremy S. McGarvey, MS (4), Gregory S. Podolej, MD, MHPE (5, 6), Colleen J. Klein, PhD, APRN, FNP-BC, FAAN (7, 8), Matthew D. Dalstrom, PhD, MPH (9), William F. Bond, MD, MS (10).

**Author Affiliation, City, and State**: (1) Clinical Intelligence and Advanced Data Lab, OSF HealthCare, Peoria, IL. (2) Digital Innovation Development, OSF HealthCare, Peoria, IL. (3) OnCall, OSF HealthCare, Peoria, IL. (4) Department of Healthcare Analytics, OSF HealthCare, Peoria, IL. (5) Department of Emergency Medicine, University of Illinois College of Medicine at Peoria, Peoria, IL. (6) OSF HealthCare, Peoria, Il. (7) Center for Advanced Practice, OSF HealthCare, Peoria, IL. (8) Saint Anthony College of Nursing, Rockford, IL. (9) Graduate Department, Saint Anthony College of Nursing, Rockford, IL. (10) Department of Emergency Medicine, Thomas Jefferson University's Sidney Kimmel College of Medicine, Philadelphia, PA.

*Note: Jonathan Handler is currently at Keylog Solutions LLC, Northbrook, IL (Keylog). During the work, he had the roles at Keylog as noted in the Declaration of Interests. However, at the time this work was performed, it was performed while he was at, performed under the auspices of, and performed in his role with, affiliation #1 above.*

**ABSTRACT**

***Background***: Emergency Department (ED) return visits are commonly reviewed for quality assurance, but are often limited (e.g., to revisits within 48-72 hours) to increase actionable finding yield while minimizing chart review burden. Those limitations may lead to missed quality improvement opportunities.

***Methods***: We conducted an exploratory, retrospective study of randomly selected ED visits to a multihospital health system having an ED revisit within 1-14 days to the same health system. Given only each visit's primary diagnosis, raters (2-3 clinicians and GPT-4 large language model [LLM]) assessed characteristics of the diagnosis pairs, including the "target": whether a pair warranted further assessment. Informed by rater response analyses, an algorithm leveraging an LLM-populated knowledge graph ("KGA") was created to automatically screen for potentially concerning pairs, then preliminarily assessed.

***Results***: 99 diagnosis pairs were included. GPT-4 responses poorly correlated to clinician raters, rating nearly all (94%) pairs as warranting follow-up (4.4-13.3 times more than clinicians). However, prompt engineering was minimal. Among clinician raters, revisit medical gravity was consistently significantly associated with the target, while a differential diagnosis/complication composite was significantly associated on unadjusted, but not adjusted (though less powered) analysis. The KGA achieved 83-100% positive predictive value for at least one clinician rater determining further assessment was warranted based on the diagnosis pair.

***Conclusion***: These results can inform next steps for improving screening with LLMs like ChatGPT. Further research is warranted to validate this preliminary work's finding that the KGA may enable enhancing the scope and yield of screening without substantially increasing reviewer workload.

## INTRODUCTION

Emergency Department (ED) return visit ("bounceback") reviews have been used for quality improvement, but have been limited by the chart review burden and low yield of actionable findings.[1–3]

Although clinical failures can be found outside a 48-72-hour revisit window,[4] that window is commonly used.[1,2,5] Due to limited resources, even more restrictive criteria may be used. However, at our highest-volume ED, even more narrow restrictions (a 48-hour return window plus revist results in admission) still leads to a large number for review and a reportedly low yield of actionable findings.

To improve clinical quality despite constrained resources, record selection for manual review must have a high positive predictive value (PPV, few false positives). Therefore, we sought to: 1) explore factors associated with determining need for revisit manual review; 2) explore automated revisit screening for manual reviews using a large language model (LLM); and 3) derive and explore an automated revisit screening algorithm ("KGA") that leverages an LLM-populated knowledge graph.

## METHODS

This work was approved by OSF Research Administration and the University of Illinois College of Medicine Peoria Institutional Review Board-1.

Encounter pairs were randomly selected from all ED encounters in a US multi-hospital health system during 2022. Only the primary encounter diagnosis was considered. Inclusion: age at index visit ("index") 19-89 years; index disposition not Admit, Send to Labor and Delvery, Transfer, Send to Ancillary Department, Deceased, or null; an ED revisit to the same health system within 1-14 days after the index visit; revisit primary ICD-10 code having differing first three characters from that of the index; not having a Z53.21 (left without being seen) primary diagnosis for either visit; and patient address mappable to a Rural-Urban Continuum Code[6] (for potential future investigation).

**Reviews**

Two reviewers were board certified emergency physicians (WFB, GSP), one (WFB) with quality improvement chart review experience. A third (JEM) was an Urgent Care Advanced Practice Provider.

Reviewers independently assessed diagnosis pairs in Microsoft Excel.[7] No other encounter-specific information was provided. For both visits in the pair, reviewers rated:

- Diagnosis Membership Value (DMV): How much of a “diagnosis” (0-100) is the diagnosis? (e.g., symptom/finding get lower scores)
- Medical Gravity Value (MGV): Given only the diagnosis, how would the reviewer respond to a patient asking, "If I have a milder form of this diagnosis (25th percentile level of badness) how bad is this diagnosis?" (0-100, worse diagnoses get higher scores). Reviewers were to enter -1 if they did not consider the diagnosis “enough of a diagnosis” for “bad” to make sense. This question was repeated for typical (median) and more severe ($75^{th}$ percentile) forms of the diagnosis.

Reviewers also rated:

- Differential Inclusion Value (DIV): The extent (0-100) to which the revisit diagnosis is in the differential of the index diagnosis, with instruction to enter -1 if the revisit diagnosis was not enough of a “diagnosis” for “in the differential” to make sense.
- Complication Inclusion Value (CIV): The extent (1-100) to which the revisit diagnosis is a complication of the index diagnosis, with instruction to enter -1 if rating did not make sense given the diagnosis pair.
- Target Variable: Whether further investigation seemed warranted given only the diagnosis pair and notation that the visits occurred within 14 days of one another.

An LLM (private Microsoft Azure OpenAI GPT-4[8] 2024-02-15-preview, or “GPT-4”) was also queried for these, each as a separate prompt. Minor prompt engineering was iteratively performed to improve response quality, so human and LLM prompt wordings were not exactly the same.

These results informed the development and preliminary evaluation (described in Results) of the KGA. The KGA depends on a knowledge graph of “potentially concerning” diagnosis pairs populated by open-source software (“Darth Vecdor”) in a previously described effort,[9] in part by repeatedly querying an LLM (GPT-4o mini[10]) about diagnoses.

**Statistical methods**

Reported preliminary descriptive analyses, analyses of rater percentage of “yes” responses for the target variable, and analyses of KGA performance were performed in Excel.[7]

Other analyses were performed using R v4.5.2[11] and assume a two-sided 5% level of significance. Values of -1 (e.g., indicating that a diagnosis was considered not enough of a diagnosis to provide a rating) were treated as missing, and records with missing values were excluded from the relevant analyses. In Stage 1 and 2 analyses (see Results below), mixed effects logistic regression was used to examine the relationship between the variables of interest and the binary target variable. Random intercepts were included for rater and study ID (for the encounter pair) to account for rater and patient-level clustering effects.

## RESULTS

Filtering criteria correction removed one case, leaving 99 encounter pairs in the dataset.

### Stage 1 Analysis (Three Human Raters)

Raters varied substantially in their responses. For the target variable (further assessment warranted), Rater 1 had 21.2% (n=21) yeses, Rater 2 had 11.1% (n=11), and Rater 3 had 7.1% (n=7). GPT-4 rated nearly all (93.9%, n=93) as yeses.

Based on investigator expertise and preliminary descriptive analysis, we chose DMV, DIV, change (delta) in "Typical" MGV, and days between visits as the following independent variables for predicting the target variable in a multivariate model. Although raters were blinded to intervisit span, it was included since greater days between visits might occur more often with clinically unrelated visits that are not judged as index clinical failures.

Only "Typical" MGV delta and DIV were statistically significantly (Table 1). Odds ratios (ORs) are per percentile change.

### Stage 2 Analysis (Two Human Raters)

We later recognized the revisit diagnosis might represent a complication of the index diagnosis and influence decisions. Two raters (WFB, JEM) assessed the extent to which the revisit diagnosis was a complication of the index diagnosis (Complication Inclusion Value [CIV]).

The investigators suspected that revisit MGV may be more relevant than MGV delta, and preliminary analyses (not reported here) demonstrated revisit and delta MGVs had relatively similar ORs with significant p-values. Therefore, additional analyses used "typical" revisit MGV rather than MGV delta.

We hypothesized that a potential index misdiagnosis (proxied by DIV) or index diagnosis complication on revisit (proxied by CIV) might influence the target and created a "relationship composite" (RC) variable having the larger of CIV or DIV for each diagnosis pair. If either DIV or CIV was -1, the non-negative value was selected. If both DIV and CIV were -1, that pair was excluded from this analysis.

RC was significantly associated with the target in unadjusted analysis (Table 2) but not adjusted analysis (Table 3). However, this may have been due to exclusions leading to a smaller sample size, since the RC unadjusted analysis no longer showed significance at a lower sample size resulting from exclusion of the same records excluded in the adjusted model (Table 3).

**Stage 3 Analysis (Algorithm Performance)**

Informed by these results, the KGA was developed, leveraging the LLM-generated KG to select pairs potentially indicative of mis-/delayed diagnosis or an index diagnosis complication (akin to the RC). The graph includes a "Medical Gravity Index" (MGI, range 44-219) for each diagnosis (analogous to the clinician-provided "MGV"). The algorithm further filtered results by revisit MGI at two cutoffs (>=90, >= 100). Returned pairs were labeled "positive" (warranting manual follow-up), all others "negative" (Figure 1).

28/99 visit pairs (28.3%) were flagged by at least one rater as warranting further investigation ("actual positives"). At the lower MGI cutoff, 5/6 were True Positives ("TPs", 83% PPV). At the higher cutoff, 4/4 were TPs (100% PPV). Among TPs, 4/5 (80%) at the lower cutoff and 4/4 (100%) at the higher cutoff were revisits at 6-7 days that would be missed by 48–72-hour revisit screening.

## LIMITATIONS

As appropriate for a brief research report, this paper had a relatively small sample size from a single (albeit large, multihospital) health system.[15] Rater agreement was also limited. Therefore, follow-up validation is needed..

Future work should explore a broader set of inputs (e.g., revisit disposition). However, these preliminary results showed high precision, and the KGA's requirement of only a few commonly available inputs may facilitate rapid implementation.

## DISCUSSION

Chart reviews for quality assessment entail high cost and effort, and computer-based automation to screen ED visits for adverse events, has been proposed to help address this.[3] Recent work used machine-learning to identify potential missed diagnostic opportunities for specific conditions.[12] LLM-based differential diagnosis generators have been studied,[13] and others have proposed using symptom-disease pairs to help identify diagnostic error.[14] Our work expands upon these efforts by incorporating all three of: 1) differential diagnosis, 2) potential complications, and 3) medical gravity. Doing so via the LLM-informed KGA aims to get LLM benefits with improved cost, speed, consistency, correctability, and explainability of results. Another strength of this work is that it was not limited to specific conditions, symptoms, or more restrictive return visit criteria.

The revisit MGV was most strongly and consistently associated with the target variable. The RC association was significant in unadjusted but not adjusted analysis. However, the latter had a smaller sample size due to more excluded cases, potentially leading to a false conclusion of insignificance. This is supported by the finding that the RC unadjusted model did not achieve significance with a dataset of only subjects included in the adjusted model.

The strengths of our reported associations are much larger than what might be inferred from their small ORs because the ORs apply to each per-point change on a ~100-point scale.

Most GPT-4 target variable responses were false positives. However, only light prompt engineering was performed, and visit pair prompts were asked separately. Asking them all at once might have allowed for a form of reasoning, potentially improving responses. Newer LLMs are now available and may perform better. Our findings suggest that a different model and/or prompt engineering would be needed to achieve better results.

The KGA shows promise as a high-precision mechanism to expand revisit analysis. Although sensitivity was low (14-18%), the appropriate comparator is the zero sensitivity for capturing cases not reviewed (e.g. revisits after 48-72 hours). For this use, the KGA's high PPV is the relevant metric.

In summary, this study explored clinician and AI-based decision-making, then used the results to inform the derivation of the KGA algorithm. These results can inform next steps for improving screening with LLMs like ChatGPT. Further research is warranted to validate this preliminary work's finding that the KGA may enable enhancing the scope (e.g., beyond the traditional 72-hour window) and yield of screening without substantially increasing reviewer workload.

**Acknowledgements**

The authors would like to gratefully thank and acknowledge Dr. Lisa Barker; Dr. John A. Vozenilek; Rebecca A. Ebert-Allen; Sarah Soorya; Claushayla M. Nunn; Milind Fougler; and Susan Wolf for their assistance in this research effort.

**Funding**

This work was funded and enabled through the generous support of OSF HealthCare Systems' OSF Innovation program.

**Data Availability Statement**

The publicly available datasets, the open-source software platform, and the platform's configurations that were all used in this study have been cited or are cited in cited work. Data used by the open-source platform to generate the knowledge graph have themselves been cited in the platform- and/or configuration-related citations. Some data from this study cannot be publicly shared because they are protected health information, subject to ethical and privacy policies and regulations, and/or subject to the limitations of the IRB and/or institutional approvals/agreements for this study.

**Declaration of Interests**

Jonathan Handler is chief executive officer and a shareholder in Keylog Solutions LLC; has received funding from Pfizer; is a shareholder in other healthcare companies including Whispersom Corporation, EmOpti LLC, HealthLab LLC, and Baxter Healthcare; is a shareholder in several companies that relate to the artificial intelligence space (e.g., Nvidia, Marvel, others); serves in an advisory role to Whispersom Corporation, EmOpti LLC, and HealthLab LLC; has various patents that have been granted or are pending; and related to activities for the respective institutions has received stipend, food, and/or travel from the American Medical Association; For the remaining authors, all declare no relevant interests to disclose. All disclosures are to the best of each author's knowledge.

**Author Contributions**

All authors attest to meeting authorship criteria as follows: Conception and Design (JAH), Interpretation of data (JAH, JSM, CJK, MDD, JEM, WFB), Drafting the article (JAH, JSM, WFB), Acquisition of data (JAH, MIR), Analysis (JAH, GSP, JEM, WFB), Revising the article critically for important intellectual content (JAH, MIR, JSM, CJK, MDD, GSP, JRM), Final approval of the version to be submitted (JAH, MIR, JSM, CJK, MDD, GSP, JEM, WFB).

**Keywords**

Quality Improvement, Quality Metrics, Artificial Intelligence, Informatics, Emergency Medicine.

**Tables**

**Table 1: Stage 1 Adjusted Logistic Regression for Warrants Follow-up**

| Independent Variable | Warrants Follow-up Odds Ratios (95% CI) (n=130) | *p* |
|---|---|---|
| Index DMV | 0.99 (0.97 – 1.01) | 0.393 |
| "Typical" MGV Delta | **1.07 (1.03 – 1.11)** | <0.001 |
| DIV | **1.02 (1.005 – 1.04)** | 0.013 |
| Days Between ED Visits | 0.96 (0.83 – 1.11) | 0.606 |

DMV: Diagnosis Membership Value; MGV: Medical Gravity Value; DIV: Differential (Diagnosis) Inclusion Value; 3 raters (WFB, GSP, JEM), 64 included unique visit pairs, 130 total observations. Records were excluded if any variables were -1 or not answered.

**Table 2. Stage 2 Unadjusted (all includable observations) Logistic Regression for Relationship Composite and Warrants Follow-up**

| Independent Variable | Warrants Follow-up Odds Ratios (95% CI) (n=180) | *p* |
|---|---|---|
| Relationship Composite (Greater of DIV or CIV) | **1.01 (1.0006 – 1.02)** | **0.038** |

Differential (Diagnosis) Inclusion Value; CIV: Complication Inclusion Value; CI: Confidence Interval. n: number of observations included. Analysis included 2 raters (WFB and JEM), 99 included unique visit pairs, 180 total observations. Records were excluded if any variables were -1 or not answered.

**Table 3. Stage 2 Adjusted and Unadjusted (including only those included in Adjusted) Logistic Regression for Warrants Follow-up**

| **Independent Variables** | **Warrants Follow-up Odds Ratios (95% CI) (n=135)** | ***p*** |
|---|---|---|
| ***Adjusted*** | | |
| Revisit "Typical" MGV | **1.08 (1.05 – 1.11)** | **<0.001** |
| Relationship Composite (Greater of DIV or CIV) | 1.01 (0.998 – 1.02) | 0.104 |
| ***Unadjusted*** | | |
| Relationship Composite (Greater of DIV or CIV) | 1.01 (0.9990 – 1.02) | 0.076 |

MGV: Medical Gravity Value; DIV: Differential (Diagnosis) Inclusion Value; CIV: Complication Inclusion Value; CI: Confidence Interval. n: number of observations included. Analysis included 2 raters (WFB and JEM), 75 included unique visit pairs, 135 total observations. Records were excluded if any variables were -1 or not answered.

**Figure 1: KG Algorithm Overview**

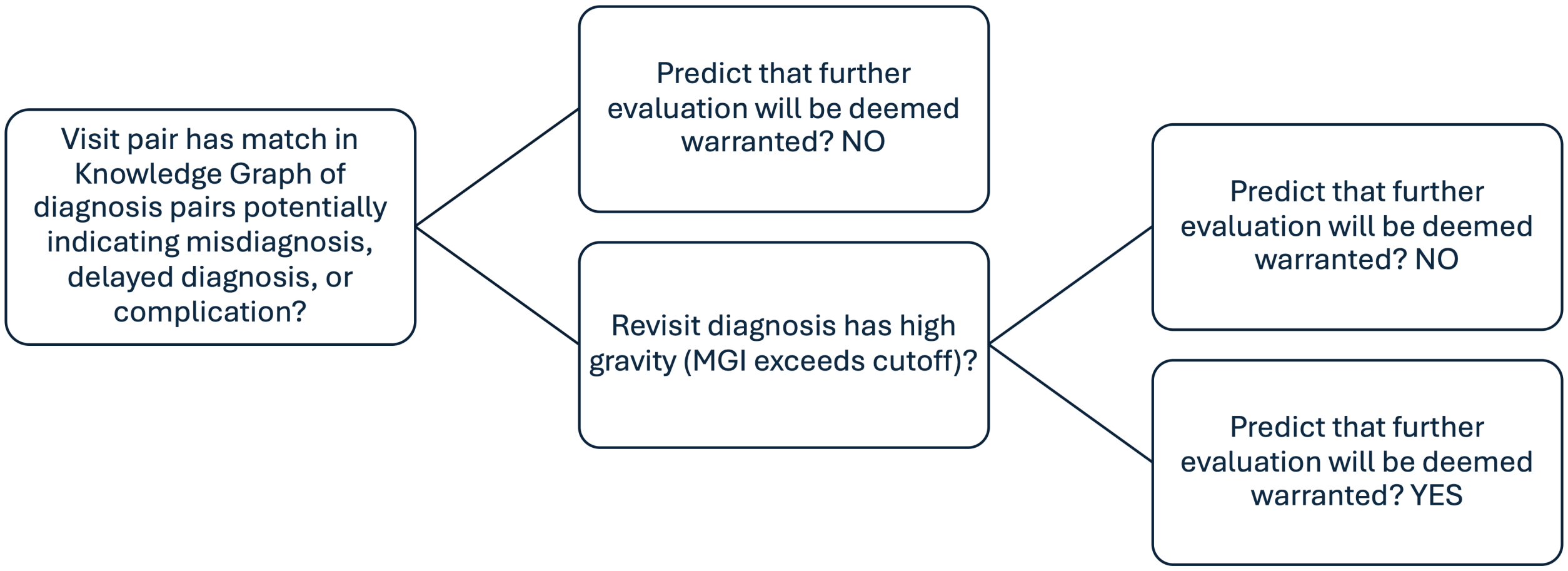

KG: Knowledge Graph; MGI: Medical Gravity Index in Knowledge Graph. As noted in the text, the Knowledge Graph was populated by open-source software, in part by repeatedly querying an LLM (GPT-4o mini) about diagnoses for the relevant relationships (e.g., complications of each diagnosis).